\documentclass[%
 reprint,
 amsmath,amssymb,
 aps,
]{revtex4-2}
\usepackage{hyperref}                         
\usepackage[all]{hypcap}                      
\usepackage{graphicx}
\usepackage{dcolumn}
\usepackage{bm}
\usepackage{xcolor}
\usepackage{caption}
\usepackage{subcaption}
\usepackage[font=small,labelfont=bf,justification=justified]{caption}

\usepackage{float}
\usepackage[english]{babel}

\begin{document}

\preprint{APS/123-QED}

\title{Influence of a static electric field on a one-dimensional Bose-Fermi mixture\\ confined in a double potential welll}

\noindent
\author{R. Avella $^1$}
\altaffiliation[]{rgavellas@unal.edu.co}
\author{J. Nisperuza$^1$, JP Rubio$^1$ and D. Grajales$^2$}\\
\affiliation{%
$^1$Fundaci\'{o}n Universitaria los Libertadores\\ Faculty of Engineering and Basic Sciences\\ Department of Aeronautical Engineering\\ A. A. 75087 Bogot\'{a}, Colombia.\\
}
\affiliation{%
$^2$ Universidad EAN\\ Faculty of Engineering \\ A. A. 110221 Bogot\'{a}, Colombia.\\
}

\date{\today}

\begin{abstract}
In this study, we conducted a detailed investigation into the time evolution of the probability density within a 1D double-well potential hosting a Bose-Fermi mixture. This system comprised spinless bosons and spin one-half fermions with weak repulsive contact interactions. Notably, even at very low effective coupling constants, periodic probabilities were observed, indicating correlated tunneling of both bosons and fermions, leading to complete miscibility, which disappears when an external electric field is turned on. The electric field accentuated fermion-fermion interactions due to the Pauli exclusion principle, altering both boson density and interactions and leading to spatial redistribution of particles. These findings underscore the complex interplay between interactions, external fields, and spatial distributions within confined quantum systems.

Our exploration of higher interaction strengths revealed conditions under which probability density functions are decoupled. Furthermore, we observed that increased fermion interaction, driven by the electric field, led to higher tunneling frequencies for both species because of the repulsive nature of the boson-fermion interaction. Conversely, increased boson-boson interaction resulted in complete tunneling of both species, especially when boson density was high, leading to effective fermion repulsion. Expanding our analysis to scenarios involving four bosons demonstrated that higher interaction values corresponded to increased oscillation frequencies in tunneling probabilities. Finally, by manipulating interaction parameters and activating the electric field, we achieved complete tunneling of both species, further increasing oscillation frequencies and resulting in intervals characterized by overlapping probability functions.

\end{abstract}



\maketitle


\section{INTRODUCTION}
The precise control of trapping geometries in experiments with ultracold gasses allows the confining of different numbers of atoms in various lattice sites using magnetic Feshbach resonance or confinement-induced resonance \cite{Ospelkaus06, Hadzibabic03, Inouye04}. By tuning both effective intra and inter-component interactions with great precision, unprecedented platforms are provided for the experimental study of quantum many-body physics \cite{Anglin02}. Various intriguing quasi-one-dimensional experiments have been conducted in harmonic \cite{Pitaevskii03, pethick_smith_2001}, double-well \cite{PhysRevLett.95.010402, Gati_2007}, periodic \cite{PhysRevLett.74.1542}, and bichromatic \cite{PhysRevA.80.023606} optical lattice traps.

Studies of dimers and single double-well systems have been conducted in one, two, and three dimensions \cite{Trotzky08, PhysRevLett.98.200405, PhysRevA.76.043606}. These systems allow for the investigation of the concept of a state as a linear superposition of 'classical' states \cite{Feynman66}, where the system can exist in a superposition of two or more degenerate states \cite{Holstein88, Chebotarev98, Garg00}. In such systems, the tunneling between local double-well potentials is negligible compared with tunneling inside the double-well potential, and each site has a well-defined and almost identical quantum state \cite{Anderlin07, Trotzky08, PhysRevA.73.033605}. This phenomenon has applications in solid-state devices, solar cells, and microscopes \cite{wiesendanger_1994}.

These systems have been used in quantum information processing, such as in the implementation of quantum logic gates \cite{PhysRevLett.98.070501}, bosonic Josephson junctions \cite{PhysRevA.55.4318, PhysRevLett.79.4950, PhysRevLett.95.010402}, squeezing and entanglement of matter waves \cite{PhysRevLett.98.030407, Esteve08}, matter wave interference \cite{ANDREWSC97, Schumm06}, and the study of exact many-body quantum dynamics in one dimension and the Josephson effect \cite{PhysRevA.79.033627, PhysRevLett.93.120401, refId0, PhysRevA.78.041403, PhysRevA.61.031601, Cataliotti01, PhysRevLett.95.010402}.

The influence of fermions on bosons has been investigated in different mixtures, such as $^4He-^3He$ \cite{Pollet08, McNamara06}, $^{87}Rb-^{40}K$ \cite{Klempt08, Karpiuk06}, $^{41}K-^{6}Li$ \cite{Lous18}, $^{87}Rb-^{40}K$ \cite{Wille06}, $^{170}Yb-^{173}Yb$, $^{174}Yb-^{173}Yb$ \cite{Sugawa11}. Other experimentally reported Bose-Fermi mixtures (BFM) include $^7Li-^6Li$ \cite{Akdeniz02}, $^{39}K-^{40}K$, and $^{41}K -^{40}K$ \cite{Vichi98}.

Bose-Fermi mixtures (BFM) have been studied numerically \cite{Roth02, Liu03, Modugno03, Adhikari04}, semi-analytically \cite{Miyakawa01, Karpiuk05}, and in the Thomas-Fermi approximation \cite{Pelster07}. The Bose-Fermi interaction induces the pairing of fermions \cite{Bijlsma00, Heiselberg00, Viverit02}, Bose phase transitions from the Mott insulator to the superfluid \cite{Mering08, Bukov14, Fehrmann04}, and  spontaneous symmetry breaking of the superfluid \cite{Adhikar10}. Other studies indicate an asymmetry between the attraction and repulsion cases \cite{Best09, Albus03}, as well as phase separation, spatial modulation \cite{Polak10}, supersolid phase, and charge density wave \cite{Titvinidze08}.

The primary objective of this research is to investigate the interaction dynamics between static electric fields and a Bose-Fermi mixture that is confined within a one-dimensional double-well potential \cite{PhysRevD.20.179, PhysRevD.20.2936, PhysRevD.21.1966}.  We consider spinless bosons in the soft-core limit and fermions possessing a spin of one-half, characterized by repulsive interactions at a temperature of absolute zero. Importantly, both bosons and fermions within the system share the same mass. Our approach leverages the two-mode approximation within the context of a double-well potential, and utilizes the lowest symmetric and antisymmetric wave functions. This particular model has demonstrated remarkable congruence with empirical findings and numerical solutions of the time-dependent Gross-Pitaevskii equation, both in one-dimensional and three-dimensional scenarios \cite{Ananikian05, Ostrovskaya00, Rey03},  and it was experimentally validated for scenarios involving weak interactions among particles \cite{Dobrzyniecki}.

The structure of this study is delineated as follows. We begin our exploration by presenting the model employed to elucidate the intricate interplay between bosonic and fermionic atoms confined in a one-dimensional double-well potential and its interaction with a static electric field in Section \ref{sec:Bose-Fermi mixtures model}. Subsequently, in Section \ref{sec:probabilities}We embark on a comprehensive examination by systematically manipulating various factors, including static electric fields and inter and intra species interacions. This dynamic manipulation was conducted to unravel the temporal evolution of probability densities within the system. Finally, in Section \ref{sec:Conclusions}, we conclude our study by offering insightful observations and remarks that encapsulate the findings and implications of our research.

\section{Physical Model}\label{sec:Bose-Fermi mixtures model}
\subsection{Double well potential}
Taking advantage of the experimental feasibility to confine quantum gasses in an array of many double-well systems, we have designed our study focusing on one of these potentials, using the experimental setup from \cite{PhysRevLett.92.050405}. In this setup, the double-well potential exhibits a radial separation of $d=13\mu m$, a trap depth of $h\times4.7Khz$, and a width of each well of $a=6\mu m$. This configuration provides confinement for isotopes of $^{170}Yb$ (bosons) and half-spin isotopes of $^{171}Yb$ (fermions). The density of fermions in the system is characterized by the parameter $\rho_F$, representing the ratio between the number of fermionic particles ($N_F$) and the number of sites ($L$) in the confinement potential ($\rho_F=\frac{N_F}{L}$), and it ranges from $0$ to $2$. In our study, we investigate two different systems. The first system involves two fermions, which results in a particle density of $\rho_F=1$, corresponding to a fermionic half-filling. In this configuration, the fermions are confined alongside a maximum of two scalar bosons $n_{max}^B=2$ \cite{Pai96, Rossini12}. In the second system, we consider a maximum of four scalar bosons $n_{max}^B=4$. These two setups will allow us to explore and compare the behavior of the particles in the double-well potential under different conditions, considering the presence of fermions and varying numbers of bosons. By analyzing these systems, we gain a comprehensive understanding of the interplay between different types of particles in a confined one-dimensional double well potential.

The system's four-particle Hamiltonian in one dimension is given by: 
\begin{equation}
\label{hubofear}
\begin{split}
\hat{H}_{BF}(x_1,x_2,x'_1,x'_2)=&\hat{H}_{B}^{\lambda_{B}}(x_1,x_2)+\hat{H}_{F}^{\lambda_{F}}(x'_1,x'_2) \\
&+\lambda_{BF}\delta(x_i-x'_j)+W^\prime(\Upsilon),
\end{split}
\end{equation}
Where $x_i$ and $x'_j$ ($i=j=1,2$) represent the space of bosons and fermions, respectively.
\begin{equation}
\begin{split}
&\hat{H}_{B}^{\lambda_{B}}(x_1,x_2)=\\
&-\frac{\hbar^{2}}{2m_{B}}\Big(\frac{d^{2}}{dx_1^{2}}+\frac{d^{2}}{dx_2^{2}}\Big)+V(x_1)+V(x_2)
+\lambda_{B}\delta(x_1-x_2),
\end{split}
\end{equation}

\begin{equation}
\begin{split}
&\hat{H}_{F}^{\lambda_{F}}(x'_1,x'_2)=\\
&-\frac{\hbar^{2}}{2m_F}\Big(\frac{d^{2}}{dx_1^{'2}}+\frac{d^{'2}}{dx_2^{'2}}\Big)+V(x'_1)+V(x'_2)
+\lambda_{F}\delta(x'_1-x'_2),
\end{split}
\end{equation}
and $W^\prime(\Upsilon)$ is the interaction energy of the  bosons and fermions placed in a static electric field. 

The system under consideration involves two different isotopes of Ytterbium, denoted as $^{170}Yb$ (bosons) and $^{171}Yb$ (fermions), with their masses represented by $m_B$ and $m_F$, respectively. Both species are subjected to the same external confinement potential $V(x)$. Due to the even nature of the function $V(x)$, we can find a basis of eigenvectors of $\hat{H}_{B(F)}$ that are either even or odd. The wave functions corresponding to these eigenvectors can be expressed as symmetrical (s) and antisymmetrical (a) linear combinations:

\begin{equation}
\label{symmetrical}
\Psi^{n}_{s(a)}(x)=\frac{\psi_i^n(x)\pm\psi_j^n(x)}{\sqrt{2}},
\end{equation}
where $\psi_{i,(j)}^n(x)$ represents a one-particle state that is twofold degenerate, and the superscript $n$ indicates the nth energy value. The states $\Psi^{n}{s}(x)$ and $\Psi^{n}{a}(x)$ correspond to the particle found in the right $|\Psi_{R}(x)\rangle$ or the left $|\Psi_{L}(x)\rangle$ of the double potential well, respectively.

The interaction between bosons (fermions) is modeled as a repulsive delta-function potential $\lambda_{B(F)}\delta(x_1^{(')}-x_2^{(')})$, where $\lambda_{B(F)}>0$. In addition, there exists a repulsive interaction between two ultracold neutral atoms of different statistics, represented by $\lambda_{BF}\delta(x_i-x'_j)$.

Because we are working with a two-mode model, both inter- and intra-particle interactions must be small enough to prevent the amplitudes of the two lower energy modes from mixing with other states. Therefore, we project the wave function onto two states, a technique commonly used in studies of Bose-Einstein condensates (BEC) in a double-well potential and Fermi superfluids. The contributions of these terms can be determined generally using the expression:

\begin{equation}
\label{int}
\lambda\iint\Psi_{i,j}^*(x_{1},x_{2})\Psi_{k,l}(x_{1},x_{2})\delta(x_{1}-x_{2})dx_{1}dx_{2},
\end{equation}

where $\Psi_{i,j}(x_{1},x_{2})$ represents the two-particle wave function and $\Psi_{i,j}^*(x_{1},x_{2})$ is its complex conjugate.

To simplify the analysis and scale the quantities, we adopt dimensionless units, where distances are measured in units of $\xi=1\mu m$ and energies are measured in units of $\varepsilon=10^{-31}J$. In these units, the dimensionless quantities are expressed as $x_a=\frac{x}{\xi}$, $E_a=\frac{E}{\varepsilon}$, and the dimensionless time as $\tau=\frac{\hbar}{\xi}$ \cite{Avella_2016}. From this point onwards, we will use these dimensionless units for lengths, energies, and time in our calculations.

\subsection{Bose and Fermi mixture in a double well}
In the previous sections, we established the necessary tools to understand the behavior of particles in a double well using the two-mode model. This model is based on the two lowest single-particle eigenstates of the external confinement potential, resulting in symmetric $|\Psi_{R}(x)\rangle$ and antisymmetric $|\Psi_{L}(x)\rangle$ states, as defined in Eq. \ref{symmetrical}. With these states, we can construct the basis for fermions and bosons by, considering the requirements of overall exchange antisymmetry and symmetry, respectively. Because the identical nature of the particles, there is a connection between their spin and the occupancy of energy levels and spatial wave functions.

For identical fermions, the subset of all possible two-particle wave functions that exhibit overall antisymmetry regarding the exchange of particle labels can be represented as linear combinations of the terms:
\begin{equation}
\Psi_s^{space}\Psi_a^{spin} \:\:  \textup{and} \:\:  \Psi_a^{space}\Psi_s^{spin}.
\end{equation}
Here, there is a singlet for $\Psi_a^{spin}$ and a triplet for $\Psi_s^{spin}$, resulting in a total of four functions.

\noindent
Singlet
\begin{equation}
\label{FerFunSin}
\frac{|\Psi_{L}(x'_1)\Psi_{R}(x'_2)\rangle-|\Psi_{R}(x'_1)\Psi_{L}(x'_2)\rangle}{\sqrt{2}}\:\:\frac{|\uparrow\downarrow\rangle-|\downarrow\uparrow\rangle}{\sqrt{2}}
 \end{equation}

\noindent
 Triplet
\begin{equation}
\label{FerFunTri}
\begin{cases}
|\Psi_{L}(x'_1)\Psi_{L}(x'_2)\rangle&|\uparrow\uparrow\rangle
 \:\:\:\:\\\
\frac{|\Psi_{L}(x'_1)\Psi_{R}(x'_2)\rangle+|\Psi_{R}(x'_1)\Psi_{L}(x'_2)\rangle}{\sqrt{2}}&\frac{|\uparrow\downarrow\rangle+|\downarrow\uparrow\rangle}{\sqrt{2}}
 \:\:\:\:\\\
|\Psi_{R}(x'_1)\Psi_{R}(x'_2)\rangle&|\downarrow\downarrow\rangle
\end{cases}
 \end{equation}

On the other hand, for bosons with spin 0, the wave function is symmetric regarding the exchange of particle labels. As a result, the wave functions are linear combinations of the terms:
\begin{equation}
\Psi_s^{space}\Psi_s^{spin} \:\:  and \:\:  \Psi_a^{space}\Psi_a^{spin}.
\end{equation}

For the specific case we are studying, there are three functions:
\begin{equation}
\label{BosFunTri}
\begin{cases}
|\Psi_{L}(x_1)\Psi_{L}(x_2)\rangle \:\:\:\:\\\
\
\frac{|\Psi_{L}(x_1)\Psi_{R}(x_2)\rangle+|\Psi_{R}(x_1)\Psi_{L}(x_2)\rangle}{\sqrt{2}}\:\:\:\:\\\
\
|\Psi_{R}(x_1)\Psi_{R}(x_2)\rangle
\end{cases}
 \end{equation}
These representations of the wave functions for fermions and bosons provide valuable insights into the behavior of particles in the double-well potential and their spin interactions.

\subsection{Static electric field}
On the basis of one-particle states $ \{\psi_{1}^1(x), \psi_{2}^1(x) \}$, the electric dipole moment can take on two opposite values, which we denote as $+\eta$ and $-\eta$. The observable associated with this physical quantity is represented as $D$, and it is described by a diagonal matrix with eigenvalues $+\eta$ and $-\eta$. The diagonal nature of the matrix indicates that the observable $D$ is well defined on this basis, and the eigenvalues correspond to the possible measurement outcomes of the electric dipole moment. Thus, the observable D for a system composed of two particles in the basis $ \{\Psi_{1,1}(x_1,x_2),\Psi_{1,2}(x_1,x_2),\Psi_{2,1}(x_1,x_2), \Psi_{2,2}(x_1,x_2)\}$ is expressed as follows

\begin{equation}
D=\begin{pmatrix}
\eta^2 &0&0&0\\
0 &-\eta^2&0&0\\
0 &0&- \eta^2& 0\\
0 & 0& 0&\eta^2\\
\end{pmatrix}
\end{equation}

When the system is placed in a static electric field $\Upsilon$, the interaction 
energy $W^\prime(\Upsilon)$ with this field is: 

\begin{equation}
W^\prime(\Upsilon)=-(W_B^\prime(\Upsilon)+W_F^\prime(\Upsilon))=-\Upsilon (D_B+D_F)
\end{equation}

For spinless bosons ($^{170}Yb$) in the Standard Model, the charge distribution is uniform and symmetric, resulting in the cancelation of any potential dipole moment. This is a consequence of the underlying symmetries of the theory, and there is no separation of positive and negative charges along any axis. In theories beyond the Standard Model, new interactions or particles could lead to nonzero electric dipole moment (EDM) for spinless bosons. However, to date, there is no widely accepted theoretical framework that predicts a non-zero electric dipole moment for spinless bosons within the context of the Standard Model. Therefore, in this study we will consider $D_B=0$.

The permanent electric dipole moment (EDM) of the $^{171}Yb$ atom is measured using atoms held in an optical dipole trap (ODT)\cite{PhysRevLett.129.083001}. They determined a value of $\eta_F=1.5\times 10 \times ^{-26} e\:cm$; there, the interaction energy of  fermions with a static electric field is considered and we focus on the subset of all possible two-particle wave functions (\ref{FerFunSin}, \ref{FerFunTri}). The interaction energy is expressed as:
\begin{equation}
W_F^\prime(\Upsilon)=\Upsilon\eta^2_F \begin{pmatrix}
1/2&-3/2&0&0\\
-3/2 &1/2&0&0\\
0 &0&- 1&0\\
0 &0&0&1\\
\end{pmatrix}
\end{equation}

\section{Probability density of the Bose-Fermi mixture.}\label{sec:probabilities}

The ground state of the system consists of two spinless soft-core bosons and two fermions with spin-1/2, positioned on the right side of the double-well potential at time $t=0$. This ground state is derived from the one-particle state (symmetric state of Eq. \ref{symmetrical}) for each species.

\begin{equation}
\label{twowave}
\begin{split}
|\Psi_{R}(x_1)\rangle|\Psi_{R}(x_2)\rangle&=\frac{1}{2}\Big[\psi_1^1(x_1)\psi_1^1(x_2)+ \psi_1^1(x_1)\psi_2^1(x_2)\\
&+\psi_2^1(x_1)\psi_1^1(x_2)+\psi_2^1(x_1)\psi_2^1(x_2)\Big],
\end{split}
\end{equation}

This configuration is primarily chosen to explore Bose-Fermi probability density as we vary the boson-boson $\lambda_{BB}$, fermion-fermion $\lambda_{FF}$, and boson-fermion $\lambda_{BF}$ repulsive contact interaction terms as well as the static electric field. 

\begin{figure}[H]
\centering
\begin{subfigure}[b]{1\linewidth}
\includegraphics[width=\linewidth]{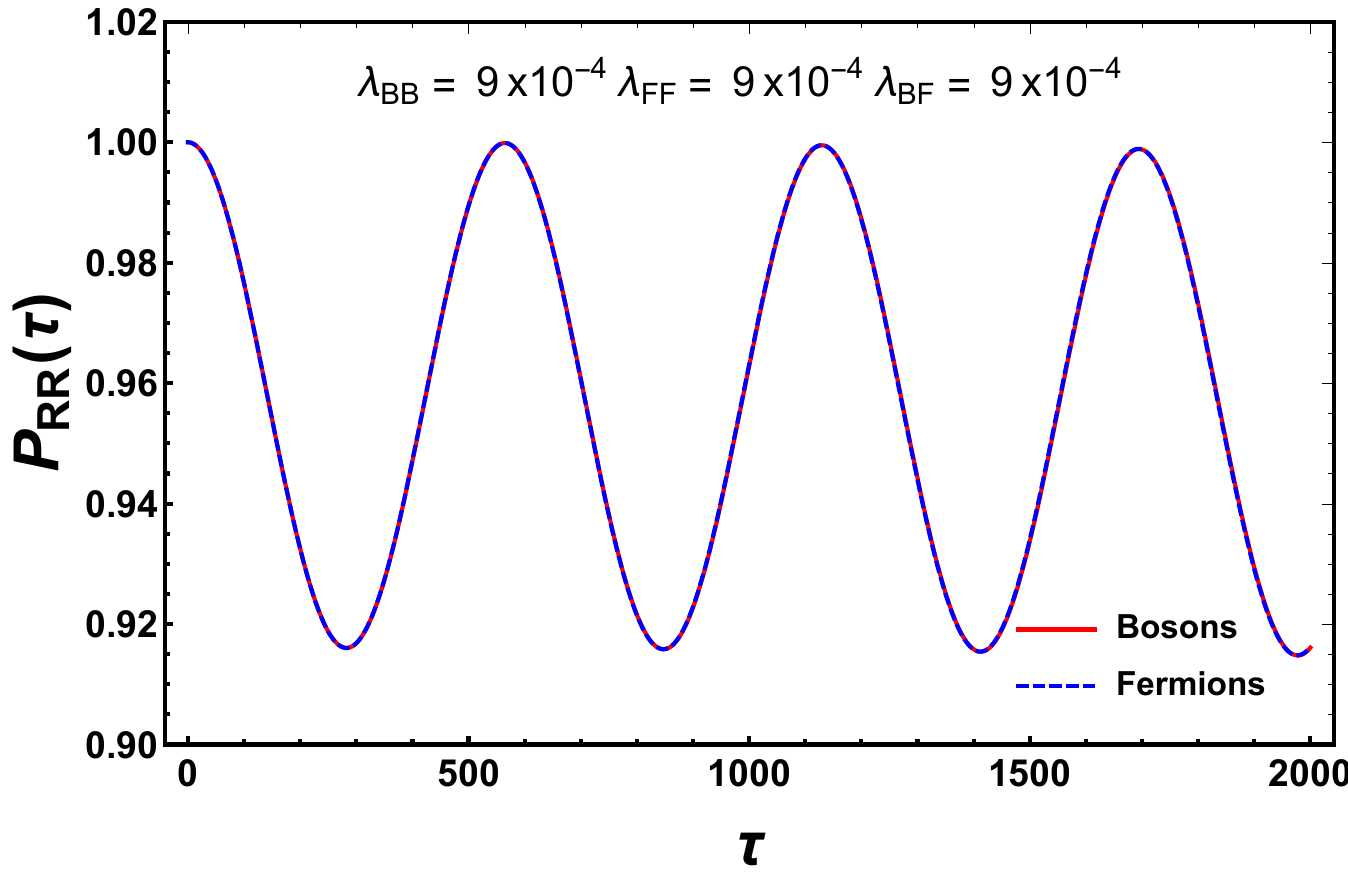}
\captionsetup{justification=justified}
\caption{When the electric field is turned, the system exhibits complete miscibility and correlated tunneling of both bosons (red line) and fermions (blue dashed line).}
\label{041}
\end{subfigure}

\begin{subfigure}[b]{1\linewidth}
\includegraphics[width=\linewidth]{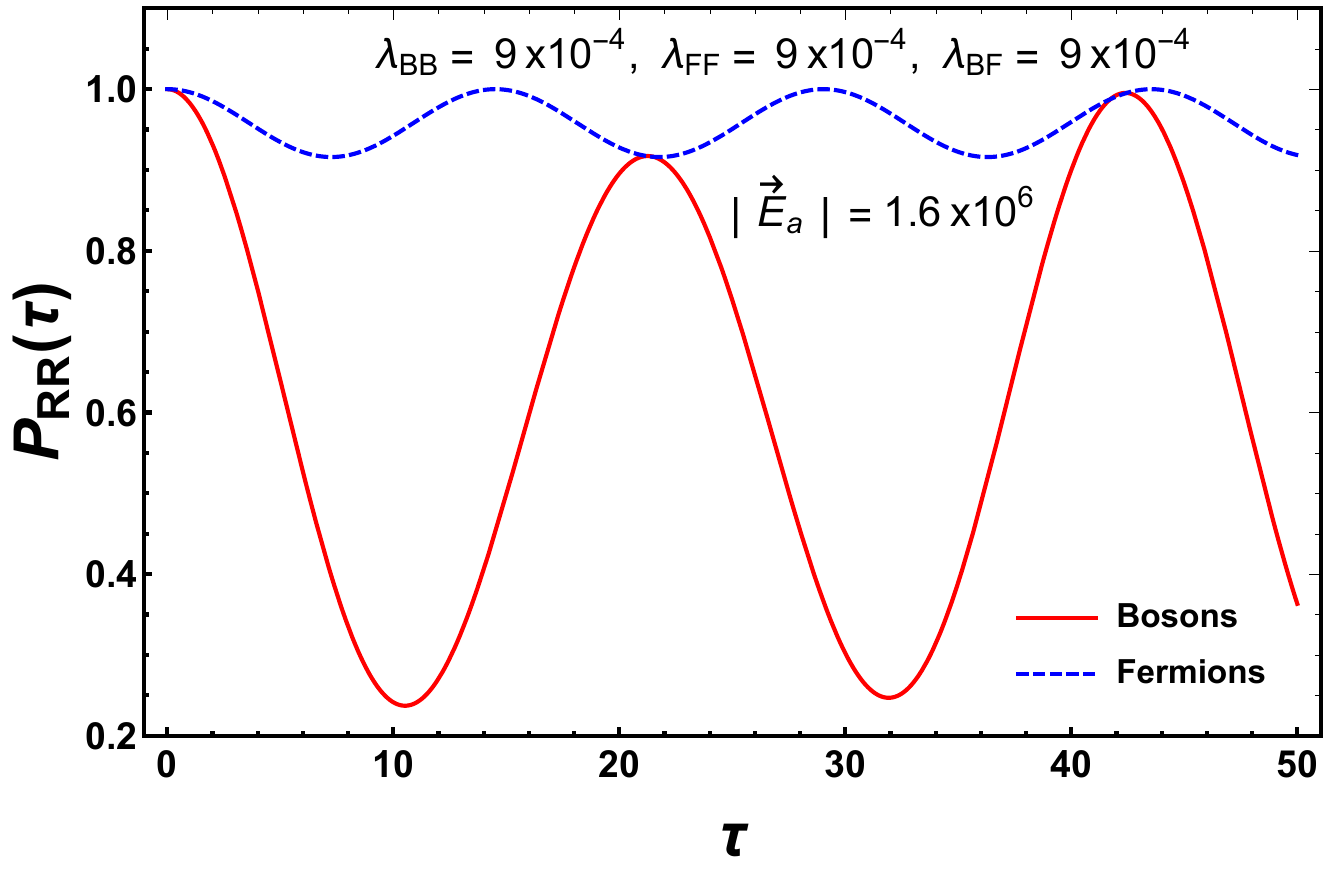}
\captionsetup{justification=justified}
\caption{The electric field is turned on ($|\vec{E}_s|=1.6\times10^6$), and the return probabilities of both species increase their tunneling frequency.}
\label{042}
\end{subfigure}

\begin{subfigure}[b]{1\linewidth}
\includegraphics[width=\linewidth]{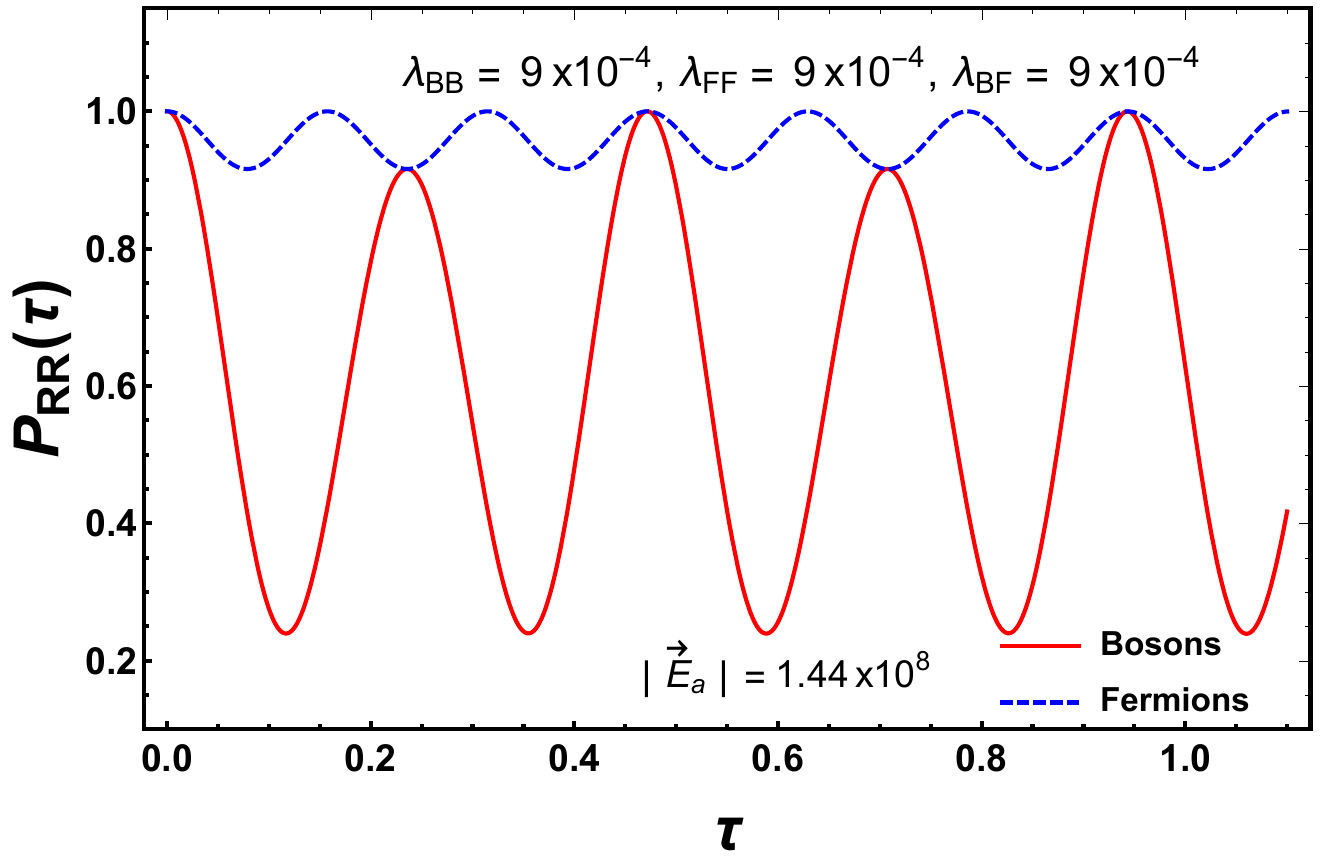}
\captionsetup{justification=justified}
\caption{The increase in the intensity of the external static electric field  ($|\vec{E}_s|=1.44\times10^8$), implies an increase in the tunneling frequency of bosons and fermions.}
\label{043}
\end{subfigure}

\caption{Time evolution of bosons and fermions probability densities on the right side of the double well $P_{RR}(\tau)$, as a function of the dimensionless parameter of time $\tau$.}
\label{symmetry}
\end{figure}

The tunneling of the quantum return probability for two particles, $P_{RR}$, for very small values of the effective coupling constants ($10^{-4}$) is characterized by periodic probabilities that exhibit correlated tunneling of both bosons \cite{Dutta15} and fermions \cite{PhysRevB.72.035330, YangYong22}, as illustrated in fig. \ref{041}. In this scenario, the bosonic probability density (represented by the red line) and the fermionic probability density (depicted by the blue dashed line) demonstrate complete miscibility \cite{jiang19, PhysRevA.97.053626}. This miscibility is defined by the overlapping of the probability densities of the two species as a function of time. This behavior persists as the values of the interaction parameters are increased while the electric field is turned off.  

When the electric field is turned on ($|\vec{E}_s|=1.6\times10^6$), the probability density functions decouple. The pair of fermions tends to remain on the right side of the double well with small fluctuations in the probability function between $1$ and $0.91$, whereas the pair of bosons increases its tunneling rate. In both cases, the frequency at which both species transition to the left side of the double well increases. 

When the bosons reach this last value, the fermions reach their minimum tunneling value, and there are times when all four particles are again on the right side of the double well,  as illustrated in fig. \ref{042}. Increasing the intensity of the electric field, the return probabilities of both species increase their tunneling frequency, exhibiting a behavior similar to that described earlier, as shown in Figure \ref{043}.

\begin{figure}[H]
\includegraphics[width=\linewidth]{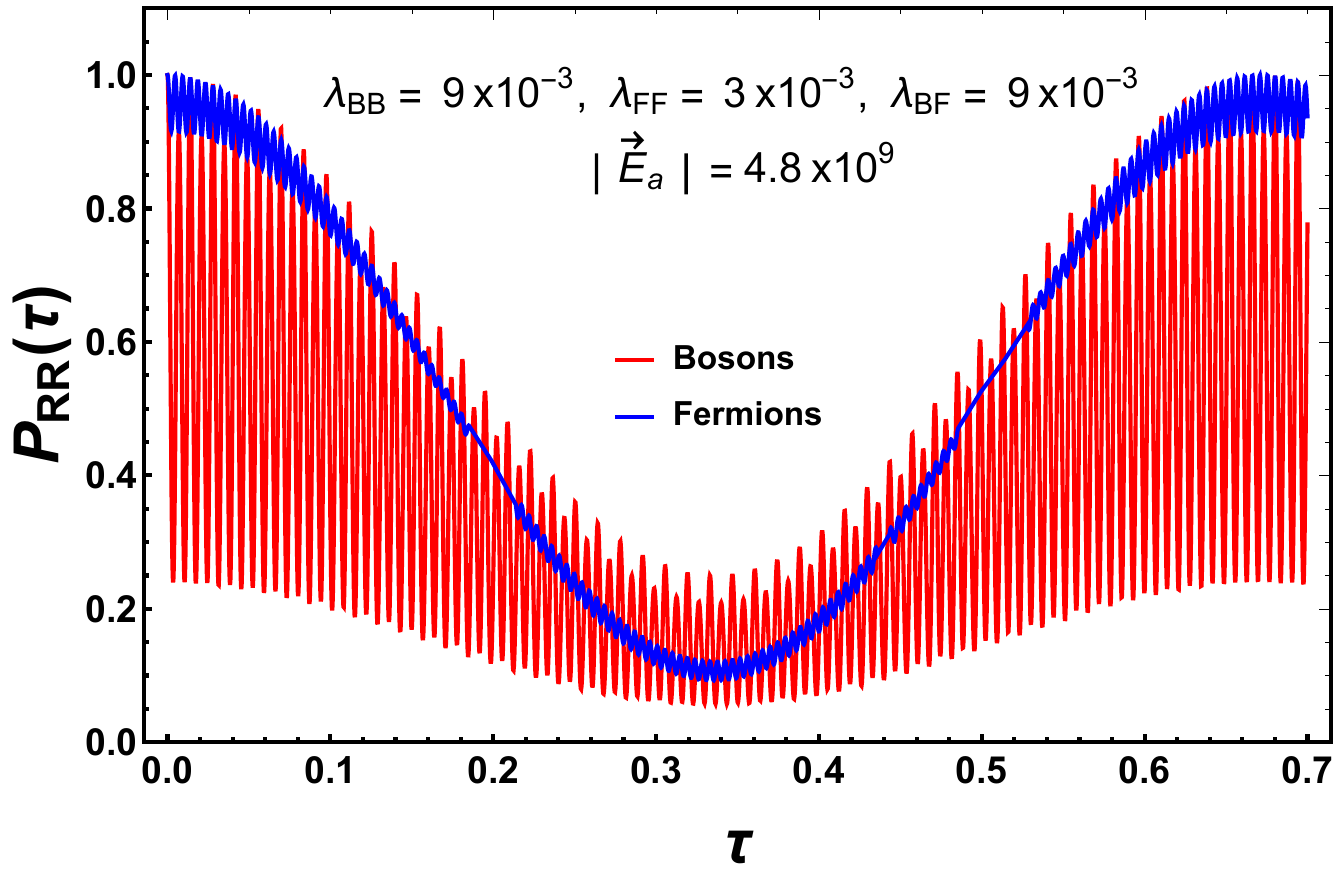}
\caption{Electric field is turned on ($|\vec{E}_s|=4.8\times10^9$). Increasing the electric field increases in the fermion-fermion interaction, thereby modifying both the boson density and their interaction due to the Pauli exclusion principle.}
\label{032}
\end{figure}

Increasing the order of magnitude of the inter- and intra-particle interactions, it is found that the probability density functions decouple when two cases are met: $\Lambda_{BB} < \Lambda_{FF}$ and $\lambda_{BB} \geq \Lambda_{BF} > \Lambda_{FF}$.  This decoupling phenomenon also occurs when turning on the electric field is turned, indicating that it increases the fermion-fermion interaction, thereby modifying both the boson density and their interaction due to the Pauli exclusion principle. Thus, as observed in Fig.  \ref{032}, in regions where the fermion density is high, bosons experience a reduction in their density. On the other hand, the variation in the boson-fermion interaction causes the particles to tend to separate spatially, promoting their spatial redistribution. Regions were also found where the density of fermions was minimal and that of bosons was maximal, as well as regions in which the density of both bosons and fermions was the same.

\begin{figure}[H]
\centering
\begin{subfigure}[b]{1\linewidth}
\includegraphics[width=\linewidth]{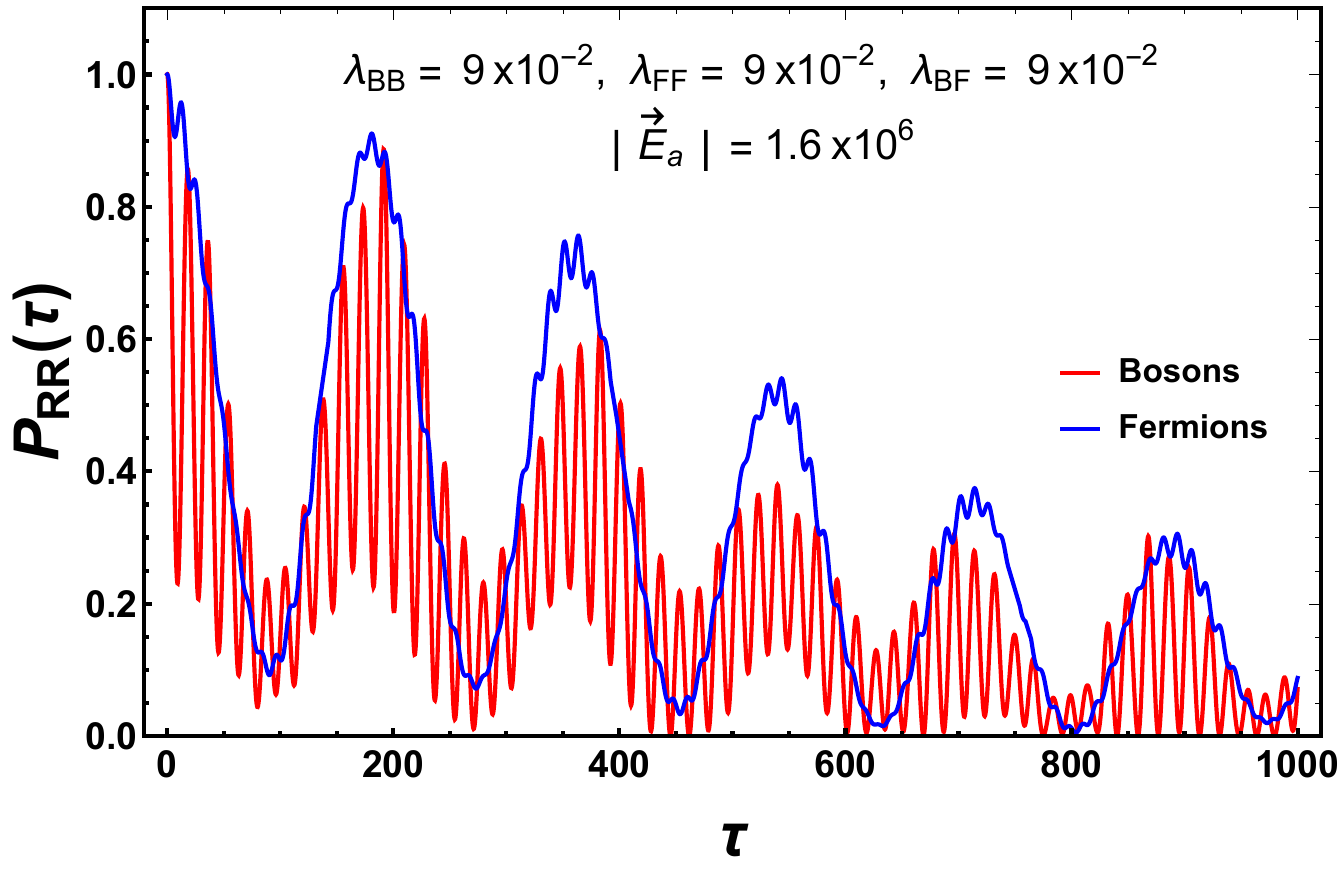}
\caption{Sequential and correlated tunnelling  of bosons and fermions through the double potential barrier as a function of the dimensionless time parameter $\tau$, for an electric field of $|\vec{E}_s|=1.6\times10^6$.}
\label{021}
\end{subfigure}

\begin{subfigure}[b]{1\linewidth}
\includegraphics[width=\linewidth]{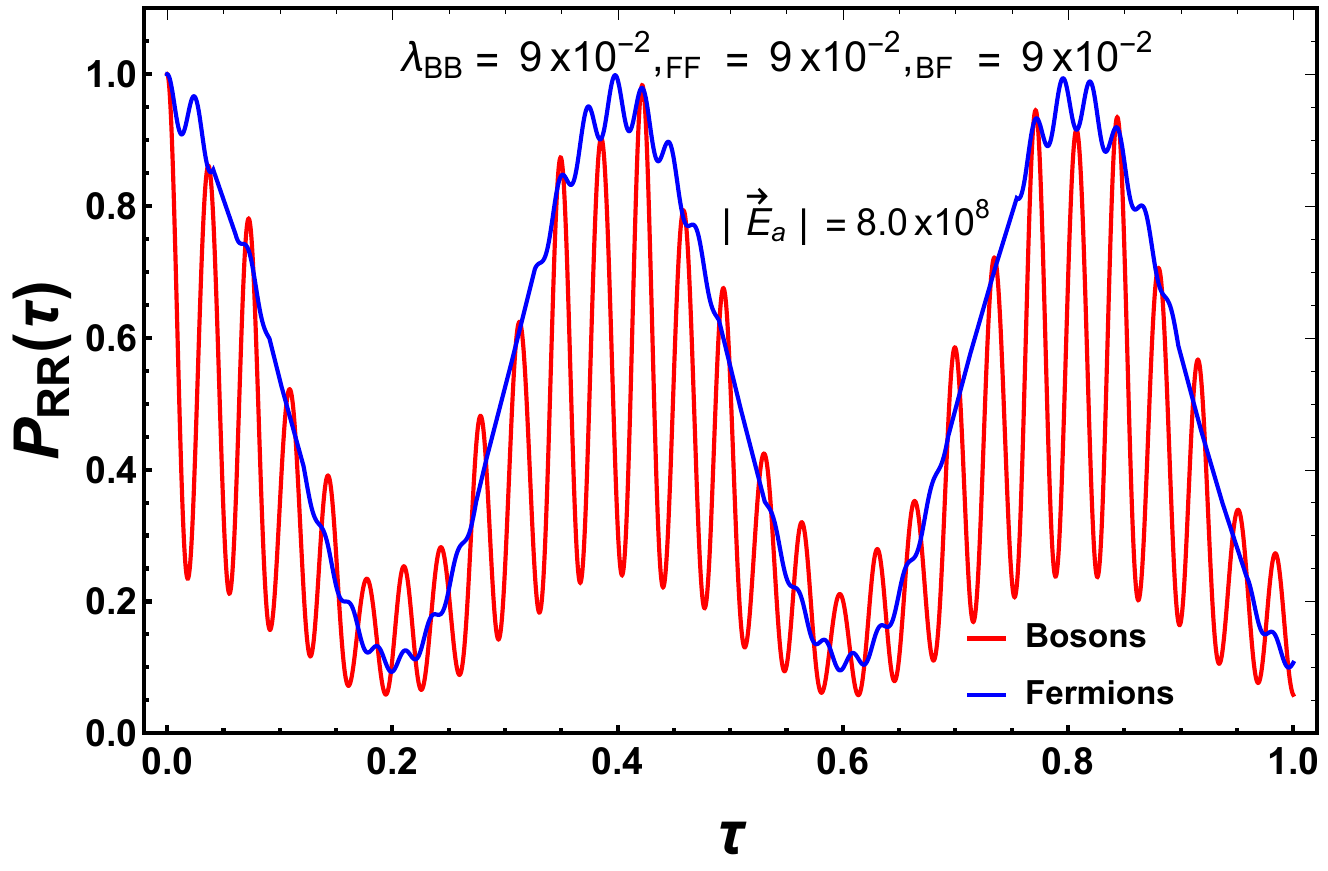}
\caption{Sequential and correlated tunneling of bosons and fermions through the double potential barrier as a function of the dimensionless time parameter $\tau$, for an electric field of  $|\vec{E}_s|=8.0\times10^8$.}
\label{022}
\end{subfigure}

\caption{Sequential and correlated tunneling between bosons (red lines) and fermions (blue dashed lines) as a function of dimensionless time for $\lambda_{BB}=\lambda_{FF}=\lambda_{BF}=9\times10^{-2}$, with electrical fields of: a) $|\vec{E}_s|=1.6\times10^6$  b) $|\vec{E}_s|=8.0\times10^8$.}
\label{02}
\end{figure}

For orders of magnitude of $10^{-2}$, we find that as the interaction between fermions increases due to the induced electric field, the tunneling frequency for both species also increases. Thus, by fixing the value of the fermion-fermion interaction, $\lambda_{FF}$, and increasing the intra-particle interaction, $\lambda_{BF}$, an increase in the tunneling frequency of both species is observed. This is primarily due to the repulsive nature of the boson-fermion interaction, which compels particles of different species to avoid occupying regions densely populated by the other, leading to a spatial redistribution of both species in the system, as depicted in Fig. \ref{02}.

To understand the influence of the boson-fermion interaction on the dynamics of fermions, the number of bosons was increased to four, and an initial state was considered in which both species were on the right side of the double well. With this configuration, the parameters of inter-species and intra-species interaction were varied, and different orders of magnitude of the external electric field were studied to investigate their influence on the dynamics of the particles confined inside the double potential well.

When considering orders of magnitude between $10^{-4}$ and $10^{-3}$, it was found that as the values of the interaction parameters increased, the oscillation frequency of the tunneling probability of the species also increased.

\begin{figure}[H]
\centering
\begin{subfigure}[b]{1\linewidth}
\includegraphics[width=\linewidth]{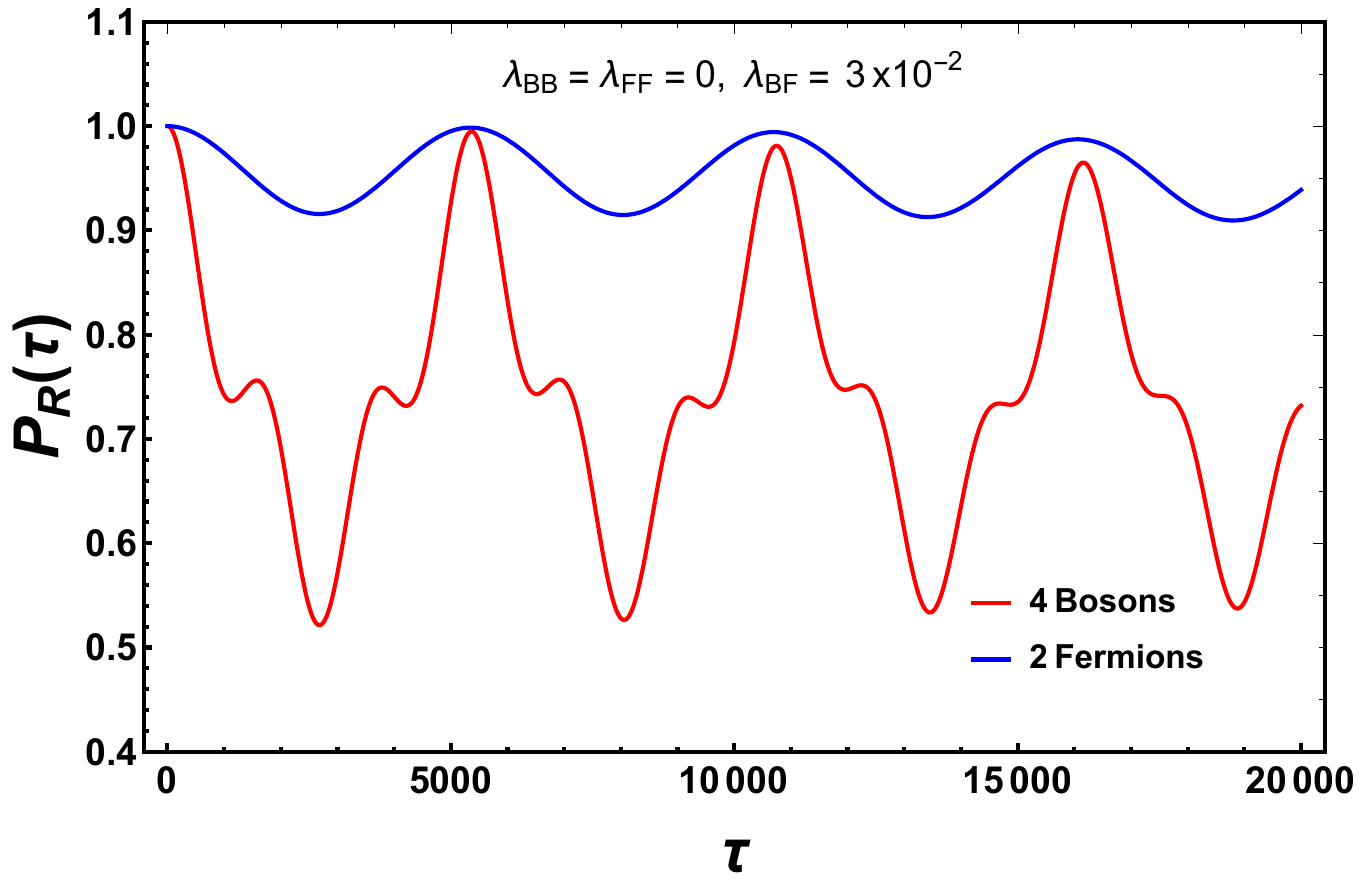}
\caption{Periodic evolution of the population of the state of bosons and fermions through the double potential barrier as a function of the dimensionless time parameter $\tau$ and  electric field of $|\vec{E}_s|=0$.}
\label{4022}
\end{subfigure}

\begin{subfigure}[b]{1\linewidth}
\includegraphics[width=\linewidth]{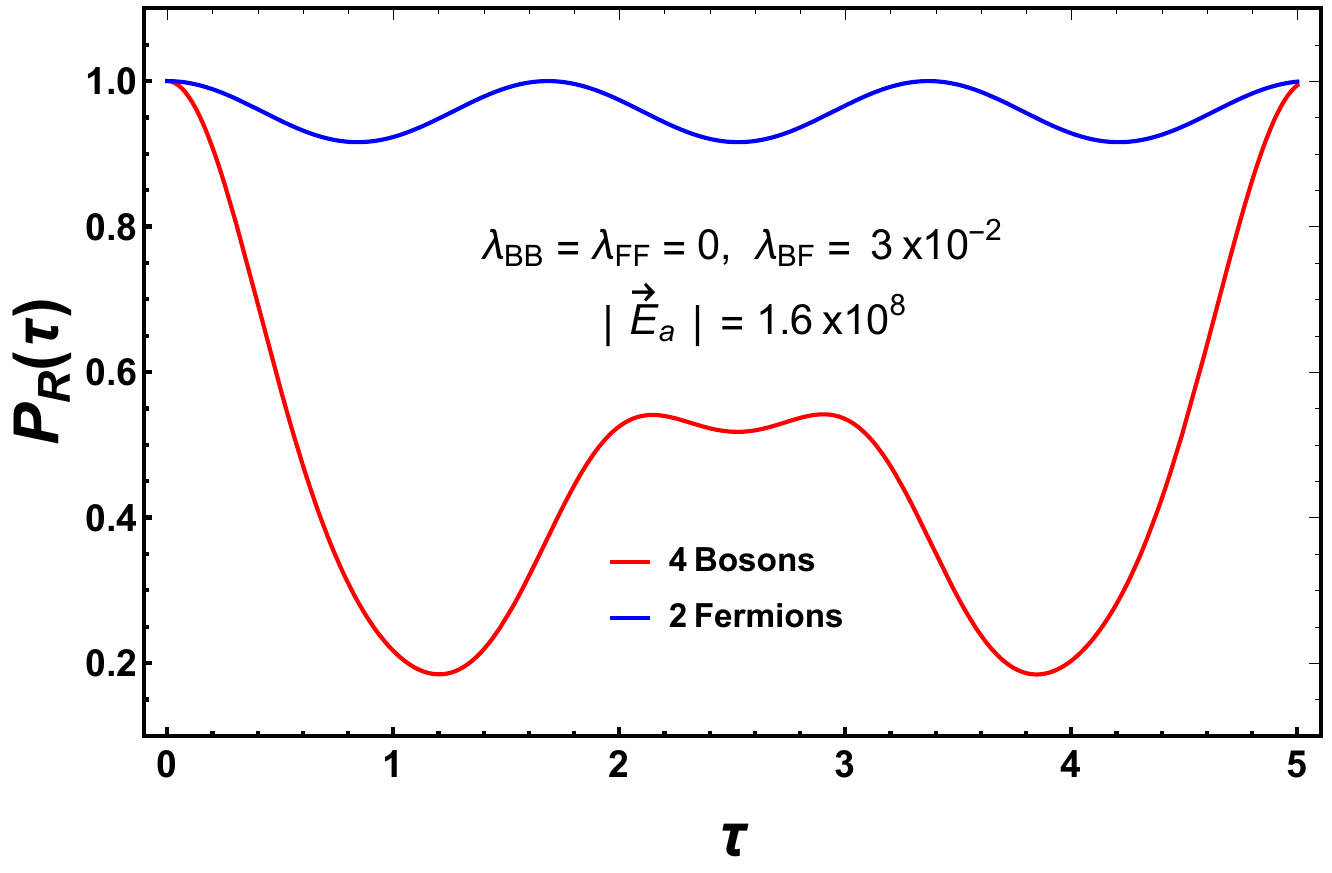}
\caption{Population state of bosons and fermions on the right side of the double potential well for $|\vec{E}_s|=1.6\times10^{8}$.}
\label{4023}
\end{subfigure}

\begin{subfigure}[b]{1\linewidth}
\includegraphics[width=\linewidth]{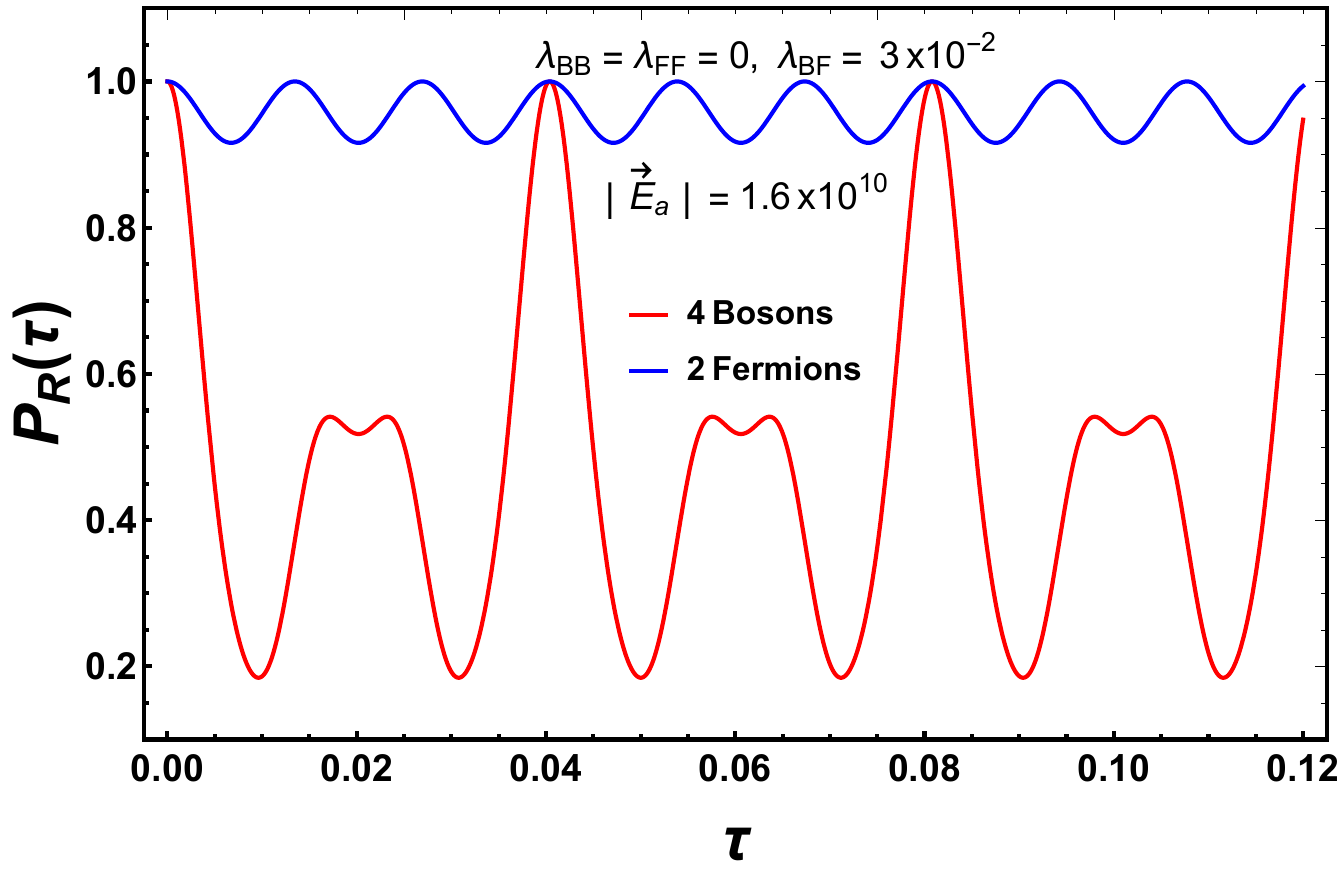}
\caption{Periodic evolution of the population state of bosons and fermions on the right side of the double potential well for $|\vec{E}_s|=1.6\times10^{10}$.}
\label{4024}
\end{subfigure}

\caption{Influence of the electric field on the probability density of bosons and fermions confined in a double-well potential and dimensionless parameters $\lambda_{BB}=\lambda_{FF}=0$, $\lambda_{BF}=3\times10^{-2}$ .}
\label{402}
\end{figure}

The study of the system, now considering interaction parameters of the order of $10^{-2}$, begins by turning off the interaction between fermions and between bosons, as well as the external electric field. With these considerations and with an interaction between bosons and fermions of $3\times10^{-2}$ (fig. \ref{4022}), it is found that fermions tend to remain on the right side of the double potential well, at least in the explored time interval. On the other hand, the temporal evolution of the probability density indicates that it is possible to find one boson at each of the ends of the double well and two in the middle region, such that $P_R(\tau) \approx0.5$. When this phenomenon occurs, the fermionic probability density reaches its minimum. It is also observed that there are time intervals where there are one and a half bosons on each side of the double potential well, indicating the presence of one boson in the center of the potential, such that $P_R(\tau) \approx0.75$.

When turning on the electric field ($|\vec{E}_s|=1.6\times10^8$), a repulsive interaction between fermions is induced, causing the oscillation frequency of the probability density to increase with a very low tunneling rate. As a result, the fermion pair tends to remain on the right side of the double well. Meanwhile, the bosons increase their tunneling rate without yet showing complete tunneling, exhibiting a time interval in which, almost constantly, two bosons are maintained at each of the ends of the double well, as observed in Fig. \ref{4023}.

When increasing the magnitude of the electric field ($|\vec{E}_s|=1.6\times10^10$), the bosonic and fermionic probability densities increase their oscillation frequency. Fermions increase their penetration rate to the left side of the double-potential well, whereas bosons exhibit time intervals in which two bosons are found at each of the ends of the double-potential well. 

There are also intervals in which all four bosons coexist on the right side of the double well, along with the two fermions, as shown in Fig. \ref{4024} increasing the magnitude of the electric field increases the oscillation frequency of the probability densities.

\begin{figure}[H]
\includegraphics[width=\linewidth]{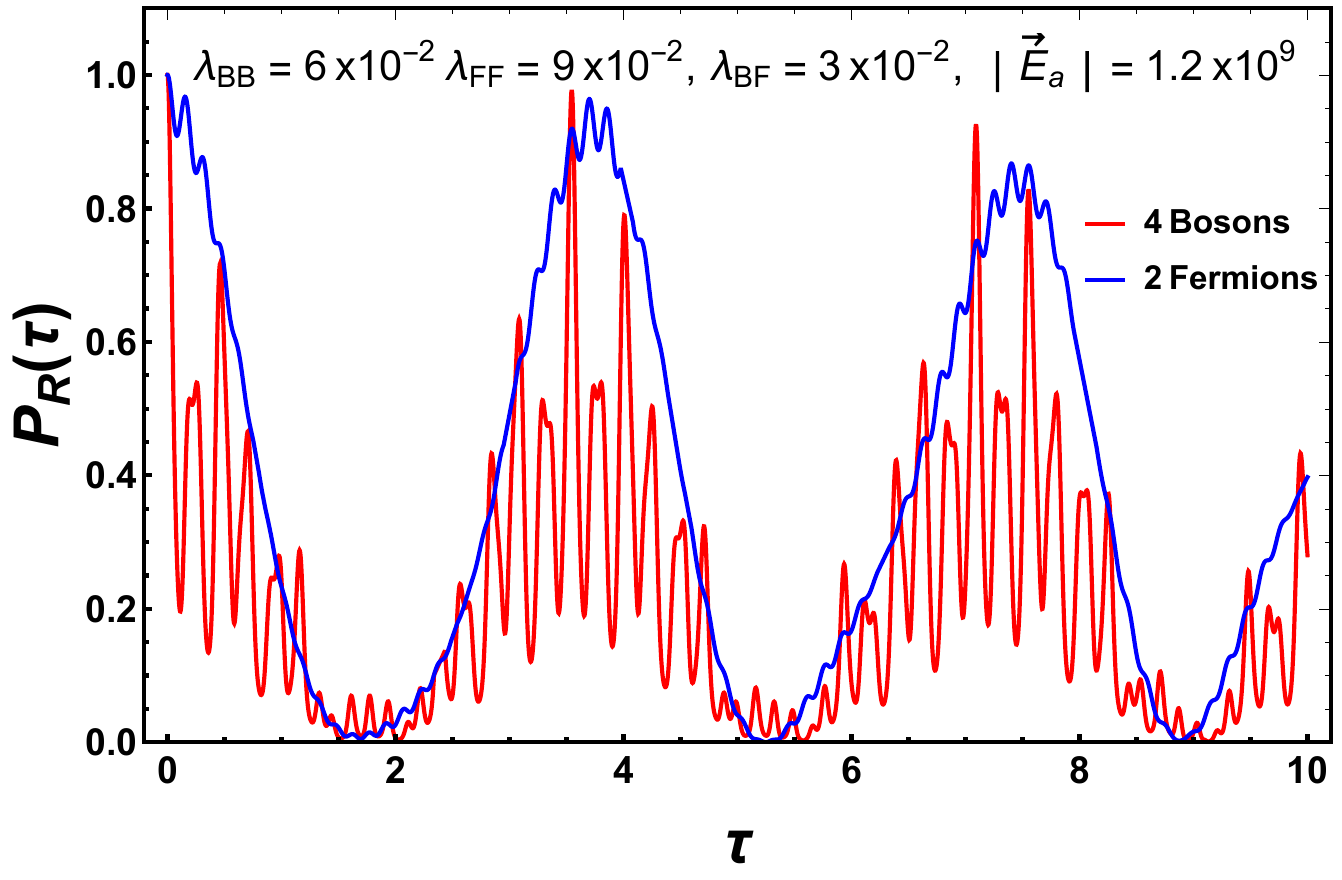}
\caption{Sequential and correlated tunneling of bosons and fermions through the double potential barrier as a function of the dimensionless time parameter $\tau$, for an electric field of  $|\vec{E}_s|=1.2\times10^9$.}
\label{4026}
\end{figure}

Thus far, we have studied the evolution of probability functions considering the interaction parameters between fermions and between bosons as zero. Because it is possible to manipulate the inter and intra-particle interactions, we varied the boson-boson interaction and found that as $\lambda_{BB}$ increases, both species tunnel completely to the left side of the double well, showing temporal intervals with overlapping of the two functions. On the other hand, when manipulating the fermion-fermion interaction, we found that the probability function of both species increases its oscillation frequency without showing complete fermion tunneling. The manipulation of both parameters, along with the activation of the electric field, allows both species to undergo complete tunneling of the double well, as well as an increase in the oscillation frequency and intervals where there is an overlap of the probability functions. The dynamics are effectively governed by the second-order process in tunneling, meaning that fermions tunnel between wells mainly in pairs (blue curve in Fig. \ref{4026}), accompanied by fully correlated tunneling of the bosons.

\section{SUMMARY AND CONCLUSION}\label{sec:Conclusions}

In conclusion, our investigation into the time evolution of probability density for a Bose-Fermi mixture in a 1D double well potential has provided valuable insights into the complex interplay of interactions and external factors affecting the tunneling behavior and spatial distribution of particles. For very small effective coupling constants, both bosons and fermions exhibit correlated tunneling, leading to complete miscibility on the right side of the double well. The introduction of an electric field decouples the probability density functions, with fermions primarily remaining on the right side of the well and bosons increasing their tunneling rate. Increasing inter- and intra-particle interaction strengths leads to the decoupling of probability density functions under specific conditions, with the Pauli exclusion principle playing a crucial role when the electric field is activated. This spatial redistribution results in regions with high fermion density, reduced boson density, and equal densities of both species. The induced electric field increasesthe tunneling frequency for both species, primarily driven by the repulsive nature of the boson-fermion interaction, leading to complex dynamics when increasing the number of bosons. Ultimately, these findings illuminate the rich behaviors that emerge in quantum systems under varying interactions and external fields, providing valuable insights into mixed quantum system dynamics within confined environments.


\section{ACKNOWLEDGMENTS}

R. A. and J.P.R. are thankful for the support of the Departamento Administrativo de Ciencia, Tecnología e Innovación (COLCIENCIAS) (Convocatoria Doctorados Nacionales 727 of 2015 and 567 of 2012 respectively.) and  gratefully acknowledge support from the "Fundación Universitaria los libertadores".

D. G. gratefully acknowledges support from " Universidad EAN".

\bibliography{apssamp}

\end{document}